\documentclass[12pt,preprint]{aastex}

\newcommand{\etal}{\mbox{et al. }}

\newcommand{\skipthis}[1]{}

\shortauthors{Trejo and Rodr\'\i guez}
\shorttitle{The Non-Thermal Radio Jet Toward NGC 2264}
\slugcomment{To appear in {\em The Astronomical Journal}}

\begin{document}



\title{The Non-thermal Radio Jet Toward the NGC 2264 Star Formation Region}

\author{Alfonso Trejo and Luis F. Rodr\'\i guez}
\affil{Centro de Radioastronom\'\i a y Astrof\'\i sica, UNAM,
Apdo. Postal 3-72, Morelia, Michoac\'an, 58089 M\'exico}

\begin{abstract}

We report sensitive VLA 3.6 cm radio observations toward the
head of the Cone nebula in NGC~2264, made 
in 2006. The purpose of these observations was to study
a non-thermal radio jet recently discovered, that appears to emanate
from the head of the Cone nebula.
The jet is highly polarized, with well-defined knots, and one-sided.
The comparison of our images with 1995 archive data indicates no evidence 
of proper motions nor polarization changes. We find reliable flux 
density variations in only one knot, which we tentatively identify as the core 
of a quasar or radio galaxy. An extragalactic location seems to be the 
best explanation for this jet.

\end{abstract}



\section{Introduction}

\begin{figure}
\epsscale{1.0}
\plotone{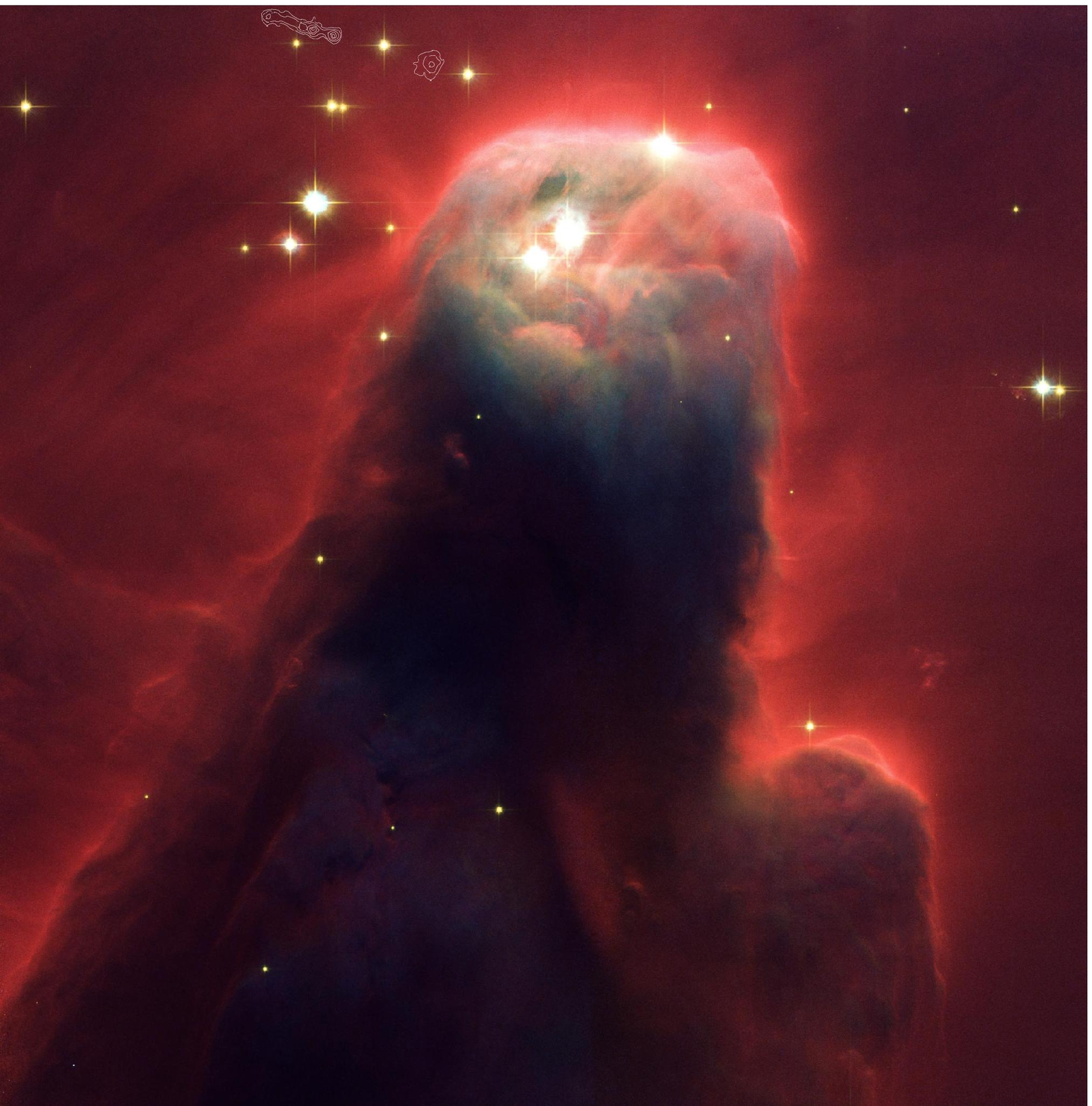}
\caption
{This optical image shows in greyscale the Cone Nebula, as observed
by the Advanced Camera for Surveys (ACS) aboard the HST.
The white contours at the top left come from our radio image of the jet (see Figure 
\ref{tres-I} for a
detailed description of these contours).
HST image Credit: NASA, Holland Ford (JHU), the ACS Science Team and ESA.
\label{cono-hst}}
\end{figure}

Reipurth \etal (2004) searched for radio continuum emission
from proto-stellar objects in eight regions of star
formation. In the NGC 2264 region they discovered a remarkably collimated 
radio jet, less than one arcminute away from the head of the well-known 
Cone Nebula and apparently emanating from it (see Fig. \ref{cono-hst}). 
Furthermore, the total flux density of the jet at 3.6 cm is $\sim$11 mJy 
and the \sl a priori \rm probability of finding a background source with 
this flux density in a region of $2' \times 2'$ is only 0.0004 (Windhorst 
\etal 1993), suggesting a possible association between the jet and the Cone 
Nebula.

The Cone nebula, discovered in 1784 by William Herschel, is believed to be 
a pillar of gas and dust whose head is most probably externally ionized by 
S Mon, a massive O-type binary (Gies \etal 1997) located about 30$'$ to
the north of the nebula. 
The head of the Cone Nebula is a dense ($\sim 10^4$ cm$^{-3}$) molecular 
core with an estimated mass of 16 $M_\odot$ (Pagani \& Nguyen-Q-Rieu 1987).
The position where the jet is located has one of the highest
surface densities of T Tau stars in the region (Dahm \& Simon 2005).  
These circumstances suggested that the radio jet could be physically 
associated with a young stellar object of the region. 
Spectacular optical jets emanating from the head of dust pillars
have been found in the Trifid (e. g. Yusef-Zadeh, Biretta, \& Wardle 2005)
and in Carina (http://antwrp.gsfc.nasa.gov/apod/ap070430.html). 

The overall extent of the jet is about 28$''$, and (with subarcsec 
resolution) it seems to be composed by eight knots.
Reipurth \etal (2004) found that assuming that the jet
was in the NGC 2264 region (at a distance of 760 pc; Sung \etal 1997)
and that the knots were moving with a velocity of 100 km s$^{-1}$ in the plane 
of the sky, an ejection every 60 years was implied. This timescale is 
consistent with that found for knots in jets associated with young stellar 
objects (e. g. Curiel \etal 1993).

On the other hand, due to the non-thermal nature of its emission and that
the jet also presents a high degree of polarization, Reipurth \etal (2004)
concluded that these characteristics were consistent with an extragalactic 
jet. The authors found no obvious counterparts in the IRAS catalogs or at 
2.2 $\mu$m emission, nor with the {\it HST} observations.
A search in the SIMBAD database shows only two stars;
V367 Mon and NGC 2264 LBM 6255 (Lamm \etal 2005) in the region of the jet, 
but neither clearly associated with any of the knots (see Fig. \ref{cono-hst}).

In this work we report an analysis of this jet, using both
new as well as archive VLA radio data. Our main goal was to compare images
taken at two different epochs to search for proper motions and variability
that could allow us to favor a galactic or an extragalactic nature for
the jet. For example, the presence of large proper motions
in the knots would favor a galactic location. We first 
searched for 3.6 cm data in the VLA archive
and found three epochs (1990, 1995 and 2002) where the
source was included in the primary beam of the
observations. These data
did not result appropriate for a reliable
search for proper motions and variability because
they have different wavelengths, pointing centers, and phase calibrators.
In order to make a reliable, high
precision comparison we made new VLA observations in 2006 that
match the parameters of the 1995 data. 
In this paper we present a comparison of these data taken with a time
separation of 10.72 years.


\section{Observations}

Our new observations were made with the NRAO\footnote{The National Radio 
Astronomy Observatory is a facility of the National Science Foundation 
operated under cooperative agreement by Associated Universities, Inc.} Very 
Large Array, in the B configuration at the wavelength of 3.6 cm, on 2006 
September 7 (we will refer to the epoch of these data, taken under
VLA project code AR599, as 2006.68).
These data have an on-source integration time of 4.9 hours. The 1995 
December 16 (epoch 1995.96 taken under
VLA project code AW420) data have, as we noted before, the same 
observational parameters and has
an on-source integration time of 3.4 hours. 
The {\it uv} coverage of both data sets is 
roughly the same, even when the 2006 data had two antennas missing at the 
center of two of the arms at the time of the observations. 

To obtain two images for reliable comparison we convolved
both images to the same angular resolution, resulting in
a beam with half power full width of $0\rlap.{''}86$. These images were made
with the ROBUST parameter of the task IMAGR set to 0.

The absolute amplitude calibration accuracy of the images is
uncertain at the $\sim$10\% level. However, due to the good uv coverage,
the relative strength of features in an individual image is measured
to higher accuracy. Thus, to allow a direct comparison of the images,
we have solved for a relative scale factor which brings the two images
into best agreement.  Specifically, we combined the images
(subtracting the 1995.96 image from the 2006.68
image) in order to get the smallest 
rms value in the difference image. The final factor for 
scaling was multiplying the 1995.96 image by 0.93.
In the following section we present and discuss these images.

To gain a better undestanding of the region
as a whole, we searched in the VLA archive for
radio continuum observations of lower angular resolution than those discussed
here. For the epoch 1984 August 31 we found observations taken at 6 cm 
under VLA project code AS204 in the 
D configuration (angular resolution of $\sim 15''$). These observations have
been published and discussed by Schwartz et al. (1985) and our conclusions,
discussed below, confirm their interpretation.
These data were reduced following the standard VLA procedures.


\section{Results and discussion}

\begin{figure}
\includegraphics[scale=0.6,angle=-90]{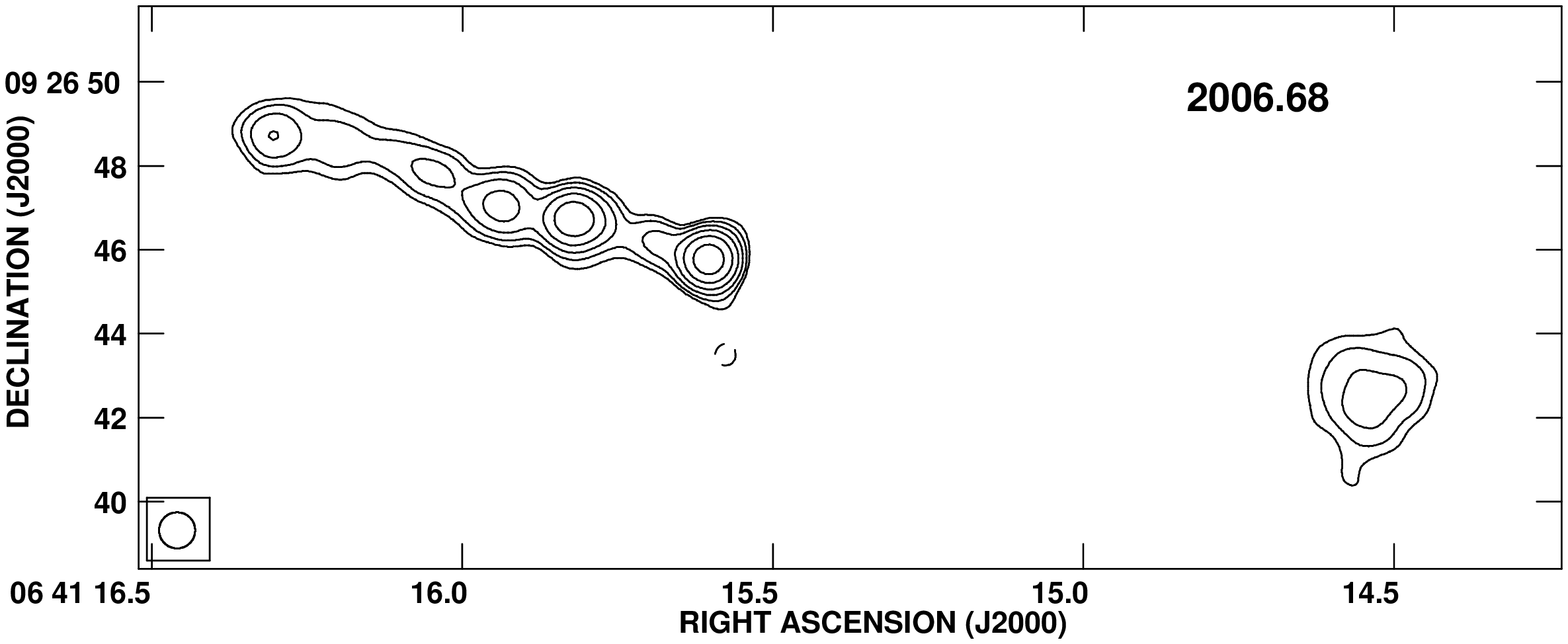}
\includegraphics[scale=0.6,angle=-90]{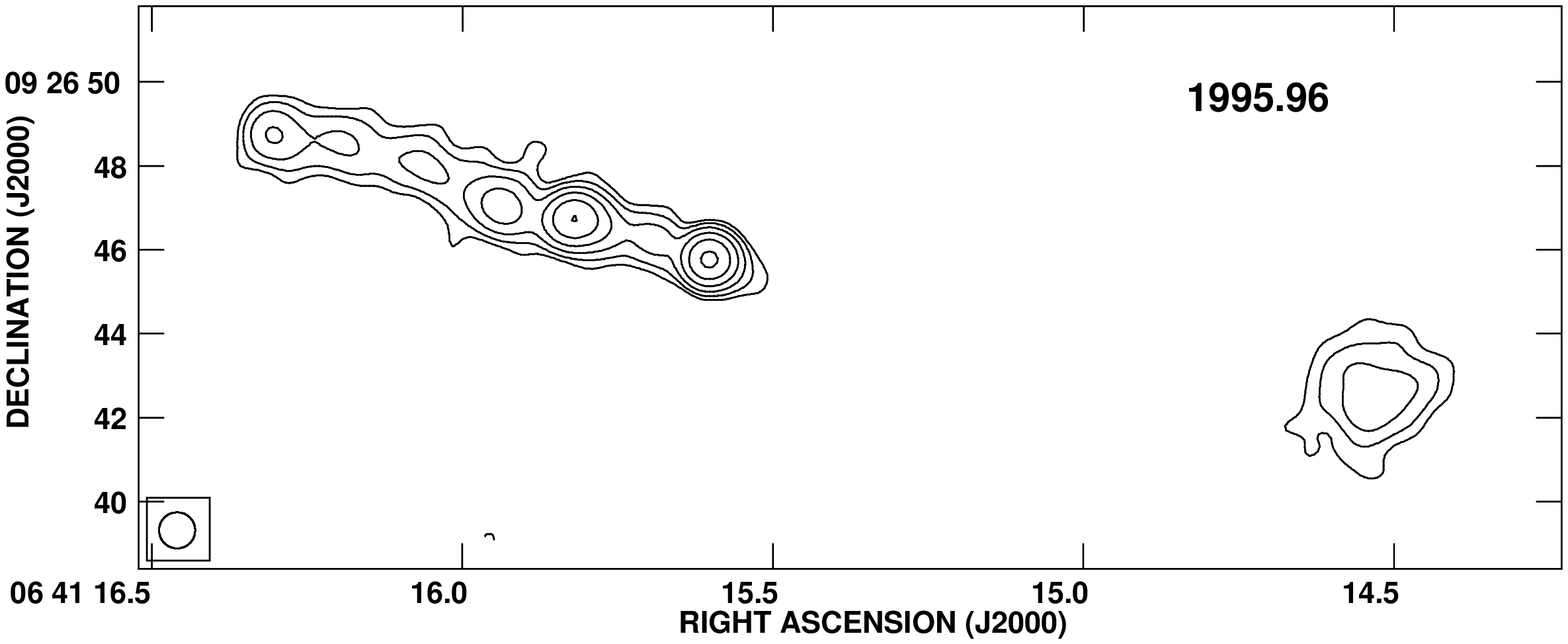}
\includegraphics[scale=0.6,angle=-90]{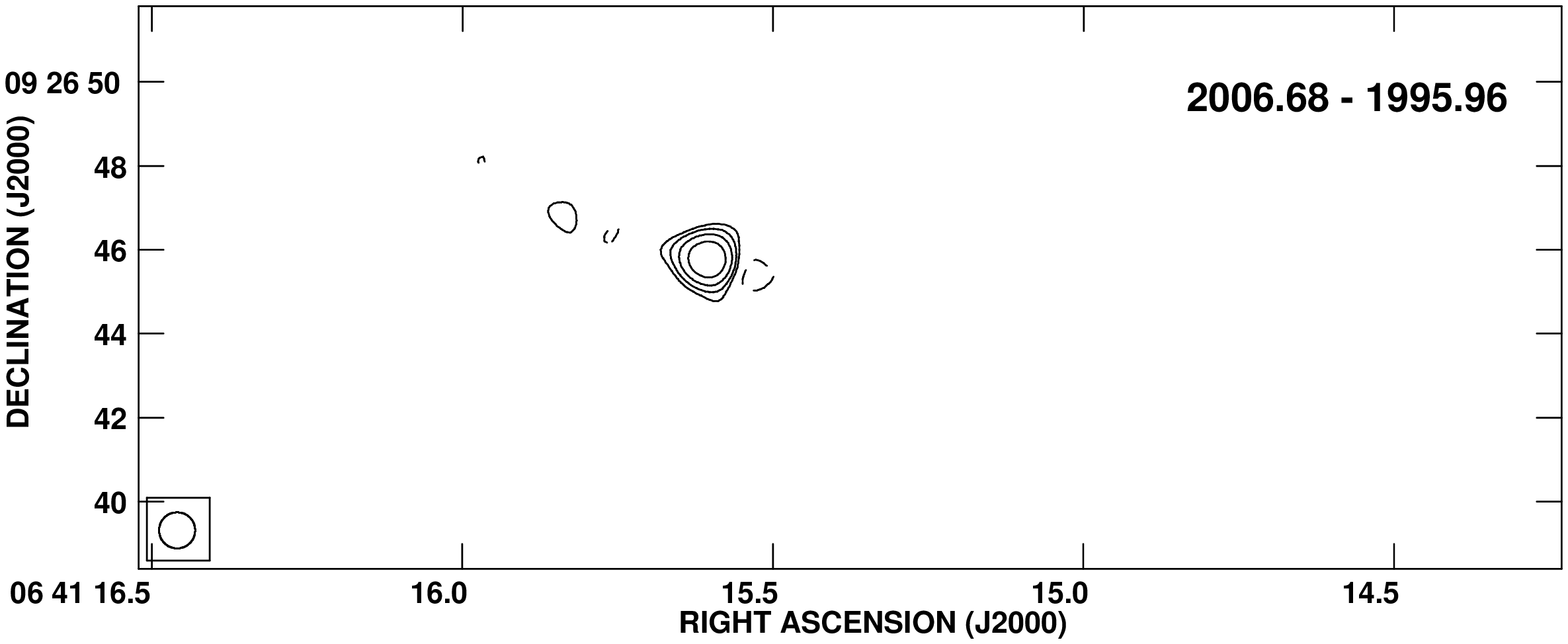}
\caption
{3.6 cm VLA images of the jet, as seen in the 2006 ({\it top}) and 1995 ({\it center)} data. 
The residual image (2006 - 1995) is shown in the bottom panel. The half-power contour 
of the restoring beam (shown in the bottom left corner) was set to
$0\rlap.{''}86 \times 0\rlap.{''}86$ ; PA = $0^\circ$ in order to make a reliable 
comparison between both epochs. The levels are -4, 4, 8, 16, 32, 64, and 128 times the rms noise 
(16.3, 13.5, and 21.1 $\mu$Jy, respectively) of each image.
\label{tres-I}}
\end{figure}

We looked for flux density and degree of polarization variations
as well as proper motions between the 1995 and 2006 images. In Figure 
\ref{tres-I} we show the continuum emission from both epochs as well as the 
difference image (2006 - 1995).
\subsection{Flux Density Variations and Search for Proper Motions}

The difference image shows no significant variation in the flux density
at a 5-$\sigma$ level of 0.1 mJy, except
for one of the knots, located at $\alpha(2000) = 06^h~ 41^m~ 15\rlap.{''}60$, 
$\delta(2000)$ = $09^\circ~ 26'~ 45\rlap.{''}774$, that we identify as the 
nucleus of the source. In contrast, none of the other knots shows variation 
above the 5-$\sigma$ level of 0.1 mJy,
that implies upper limits to any variability
in the range of 3 to 15 \%, according to their flux density. 

We also searched for proper motions in the knots, setting typical 3-$\sigma$ 
upper limits of 1.6 mas yr$^{-1}$. If the source was located at the distance 
of NGC 2264, these upper limits would imply 3-$\sigma$ upper limits of 5.8 
km s$^{-1}$ for the motion in the plane of the sky. Knots in outflows 
associated with star formation regions are known to move at velocities in 
the range of 100 to 500 km s$^{-1}$ (Rodr\'\i guez \etal 1989; Curiel \etal 
2006). In the case of galactic microquasars (Mirabel \& Rodr\'\i guez 1999),
the velocities  are much larger, comparable to the speed of light.
We conclude that the observed lack of flux density variations 
and in particular the stringent upper limits to any proper motions
argue strongly against a galactic nature for this jet.

As mentioned before, the only knot that is found to be variable in flux density
most probably traces the nucleus of this extragalactic jet. This component 
increased its flux density from 2.3$\pm$0.1 to 3.7$\pm$0.1 mJy over the period of the 
observations. 

\subsection{Linear Polarization}

\begin{figure}
\epsscale{1.0}
\plotone{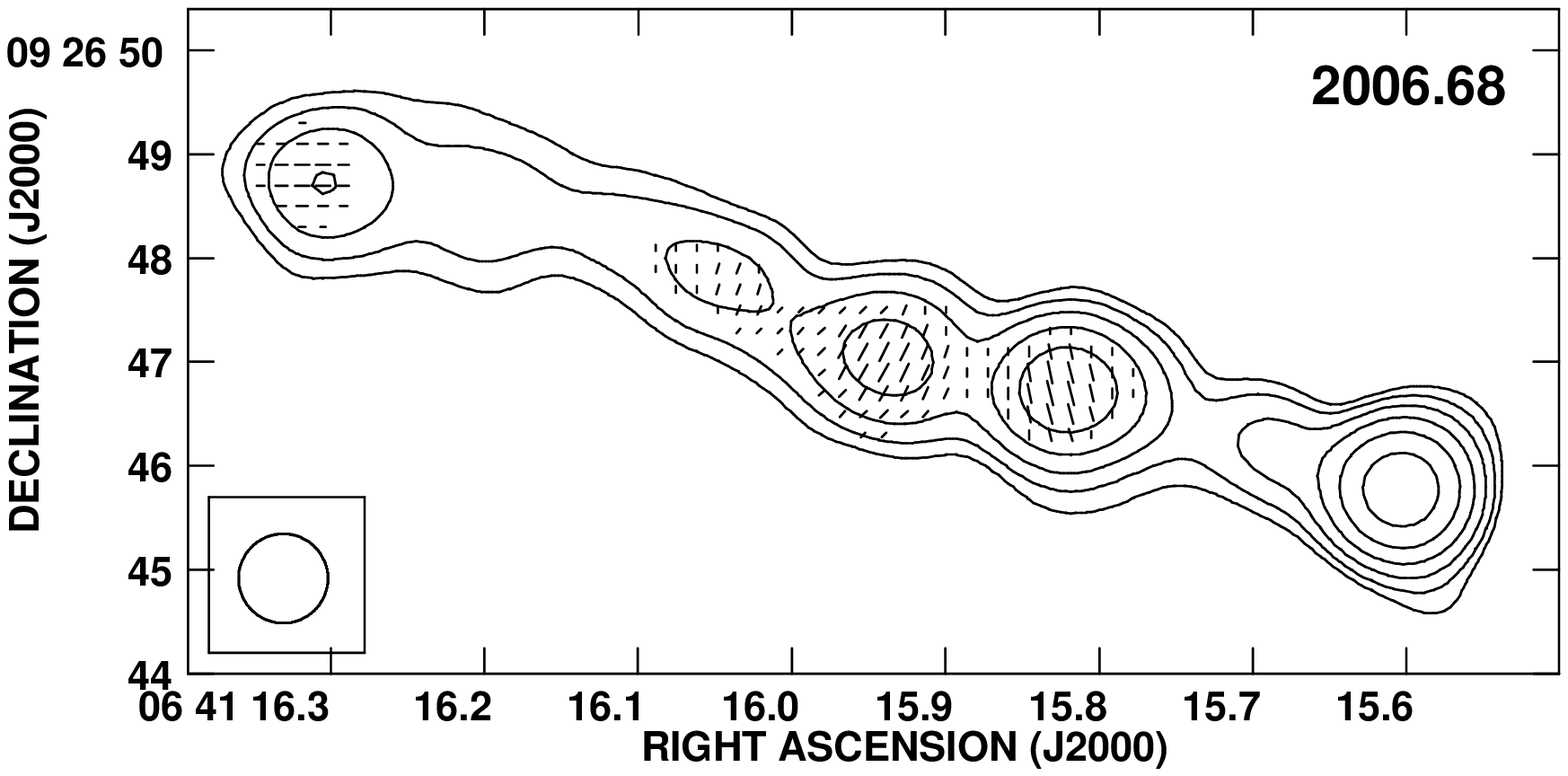}
\plotone{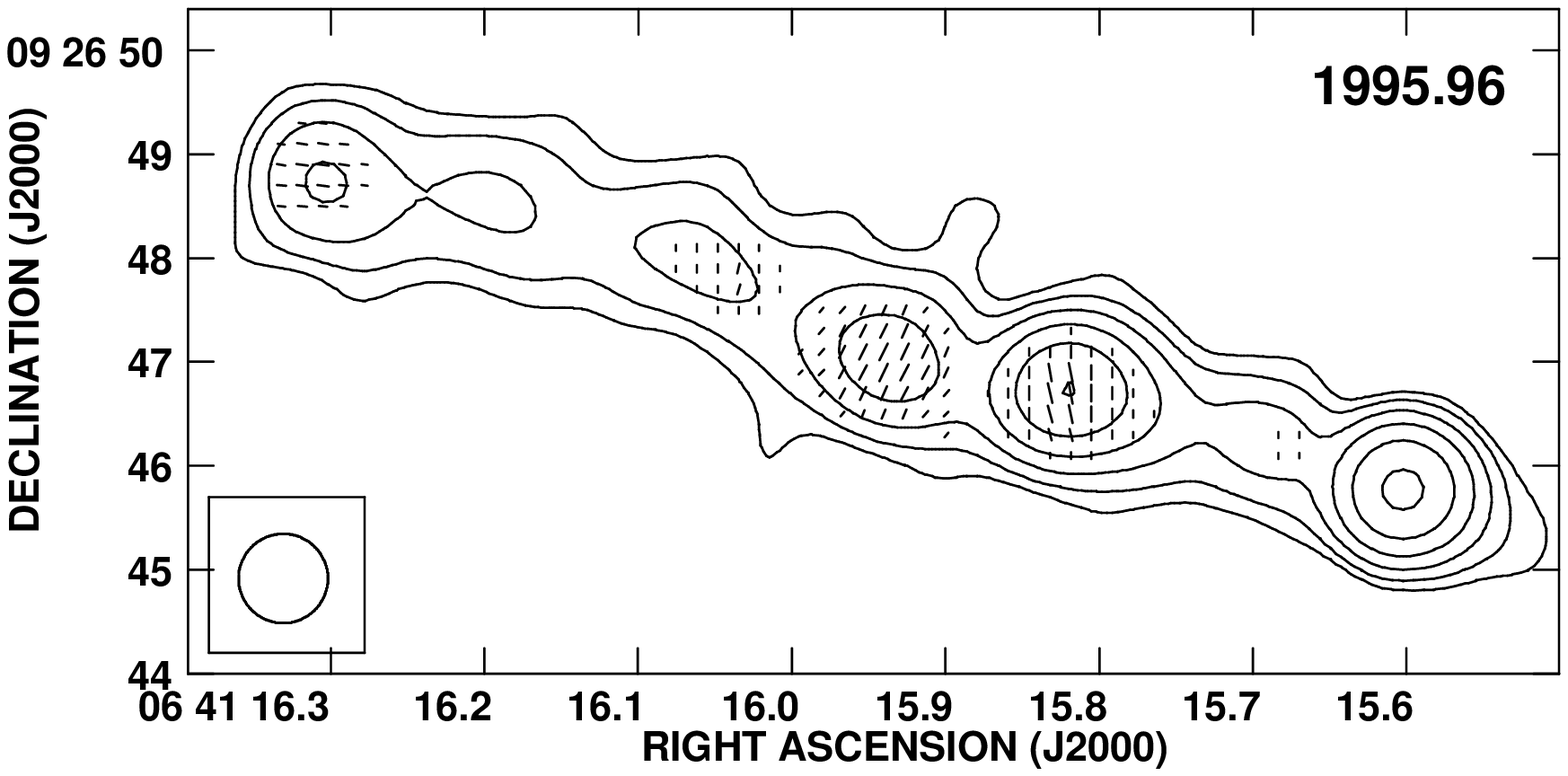}
\caption
{VLA images showing the polarization of the east side of the jet for 
2006 ({\it top}) and 1995 ({\it bottom}). The contours are the same as in Figure 
\ref{tres-I}. The 
vectors represent the electric vector and its length is proportional to the degree of 
polarization (0.20 arcsec is 25\% linear polarization).
\label{dos-P1}}
\end{figure}

\begin{figure}
\epsscale{1.0}
\plottwo{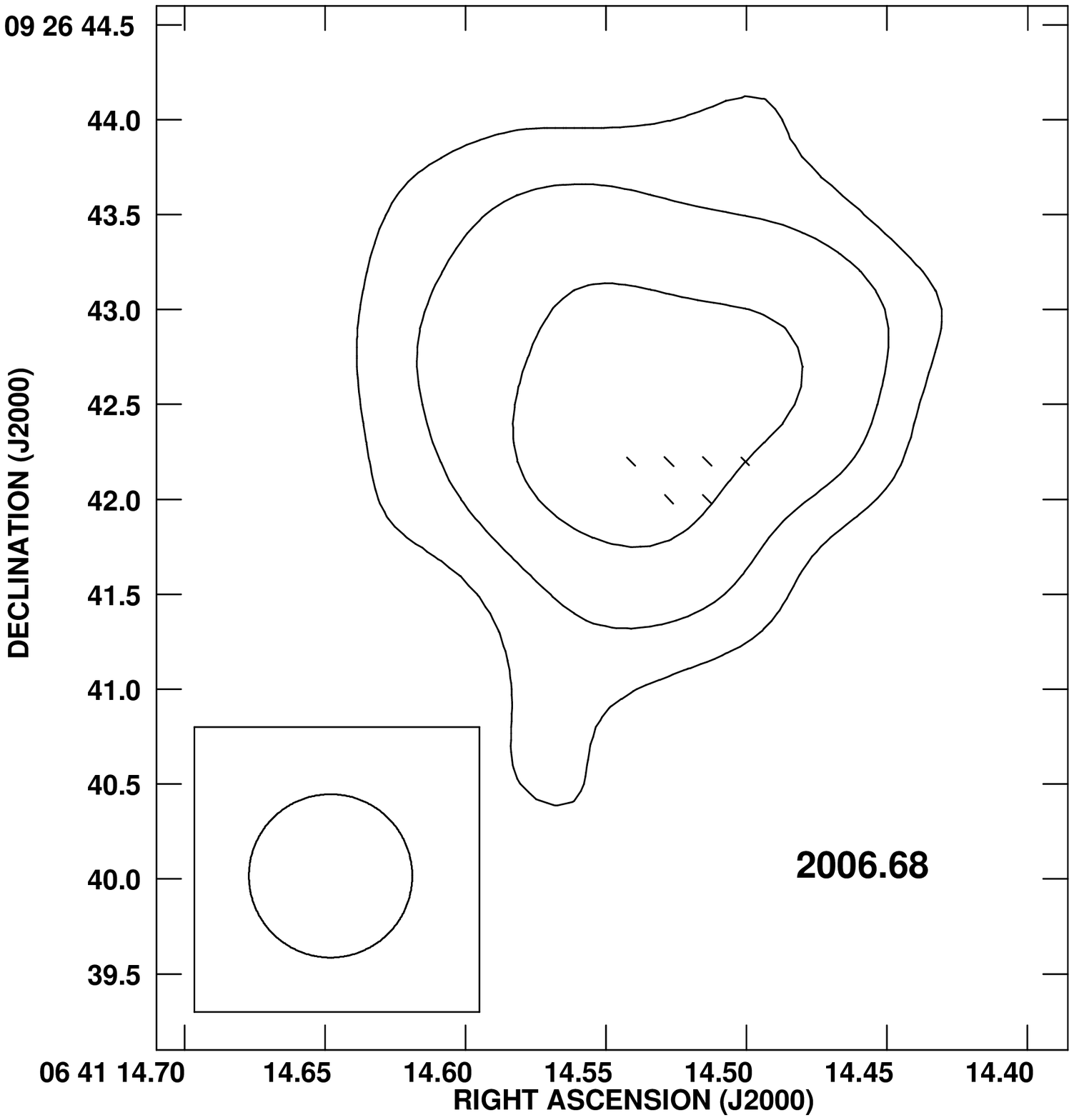}{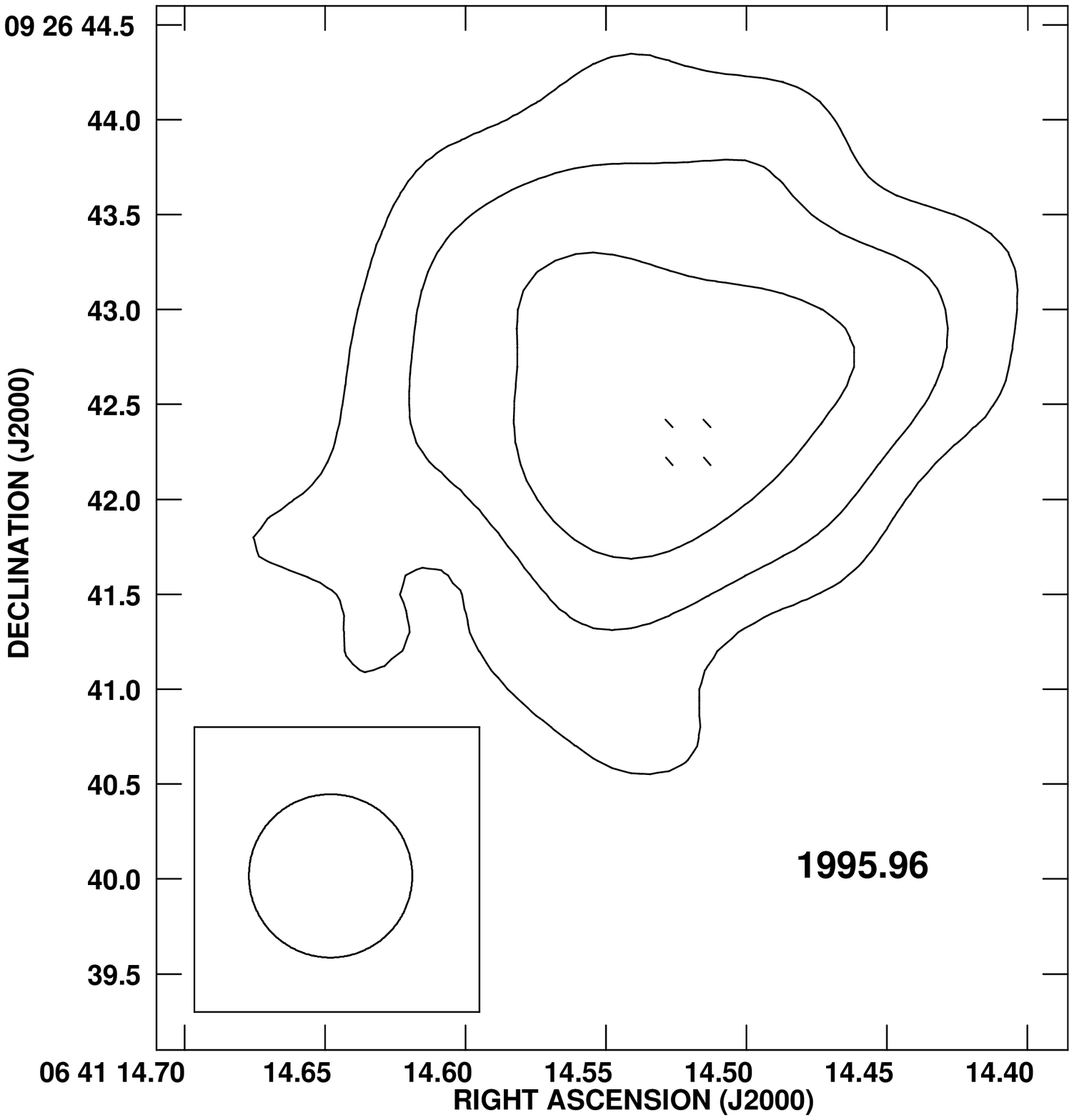}
\caption
{VLA images showing the polarization of the west side of the jet for 
2006 ({\it left}) and 1995 ({\it right}). The parameters are as in Figure \ref{dos-P1}.
\label{dos-P2}}
\end{figure}

On the other hand, the jet presents highly polarized emission, in some knots
up to 30\% (see Figure \ref{dos-P1}), including the single knot (most probably 
a lobe) seen to the west (see Figure \ref{dos-P2}).
One important fact is the absence of detectable polarized emission in the knot 
presenting variations in flux density (see Figure \ref{dos-P1}). This supports 
the idea that this knot is probably the core of a quasar or radio galaxy, that 
are known to be rapidly variable and to show a small degree of linear 
polarization, of order 1\% (Saikia \& Kulkarni 1998). Figure \ref{dos-P1} shows 
that the magnetic field is almost parallel to the jet along most of its extension, 
except in the final knot where it turns to be almost perpendicular. This 
polarization behaviour is similar in orientation and percentage to that observed 
in the jet of the well-known quasar 3C 273 (Conway \etal 1993). 
Due to the ratio of the fluxes of both sides of the jet, we classified this jet as
being one-sided, applying the rules given by Bridle \& Perley (1984) and 
Fanaroff \& Riley (1974). This jet seems to be also of type FR II, but the 
classification implies bright hot spots in the outer regions, which are not 
present in this jet.

\subsection{Large Scale Radio Emission}

\begin{figure}
\epsscale{1.0}
\plotone{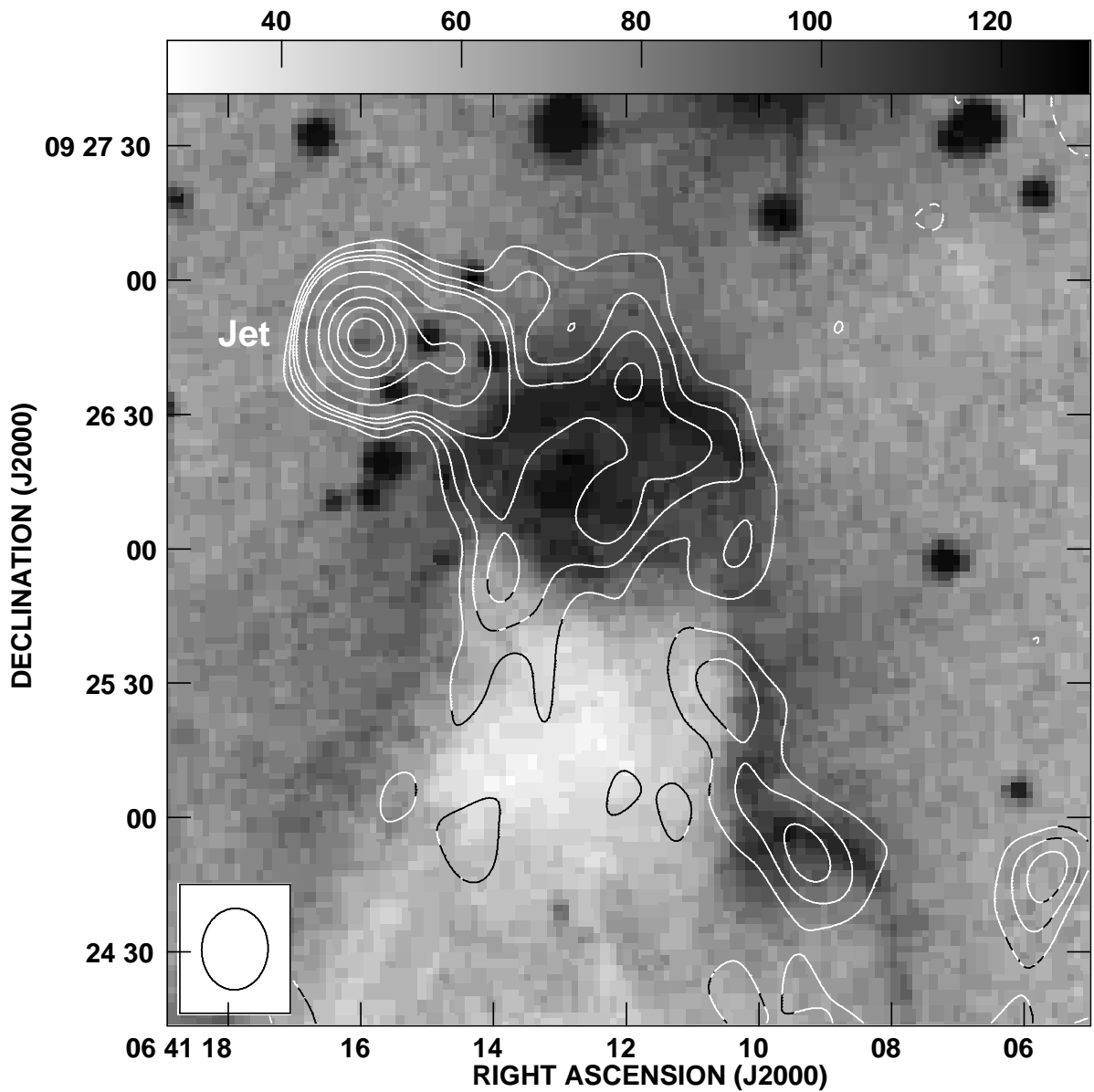}
\vspace{-5.0mm}
\caption
{This figure shows the Cone nebula in greyscale taken from the red image
of the DSS archives. The greyscale bar at the top indicates the 
intensity of the optical image in arbitrary units. The 
contours comes from 6 cm VLA-D archive data. The bright, compact double-source
to the NE of the head is the non thermal radio jet studied in this paper. 
The diffuse radio emission associated with the head and ``shoulder'' of the Cone
nebula is most probably free-free emission from gas photoionized by the massive O-type binary S Mon,
located about 30$'$ to
the north of the nebula. 
The contours are -4, 4, 6, 8, 10, 20, 40, 60, 80, and 100 times 112.7 $\mu$ Jy, the rms noise 
of the radio image. The half-power contour of the synthesized beam of the radio image
($18\rlap.{''}5 \times 15\rlap.{''}0; PA = 0^\circ$) is shown 
in the bottom left corner.
\label{cono-dss}}
\end{figure}

An image of the VLA archive data 
is shown in Figure \ref{cono-dss}. This image is superposed on the STScI Digitized
Sky Survey (DSS) red image for the same region.
From this overlay, we can see that the Cone nebula is associated with two
types of radio emission. The compact double source to its NE is the radio jet
discussed here. The more diffuse radio emission associated with the optical
emission from the head and the 
``shoulder'' of the Cone is most likely free-free emission from gas 
photoionized by S Mon, the massive O-type binary located about 30$'$ to
the north of the nebula. 
This diffuse radio emission has a total flux density of $\sim$10 mJy.
If we assume that it is coming from optically thin free-free emission
at an electron temperature of $10^4$ K, and that the region is located
at a distance of 760 pc, an ionizing photon rate of
$5 \times 10^{44}$ $s^{-1}$ is required. Furthermore, assuming
that the region of diffuse free-free emission subtends an angular
diameter of 1$'$ with respect to the angular separation of 30$'$
from S Mon, a solid angle correction
factor of $1.4 \times 10^4$ gives a total ionizing photon rate of
$7 \times 10^{48}$ $s^{-1}$ for S Mon. This is consistent with the
rate expected from an O7V star, as S Mon is classified (Pagani 1973). 
We then conclude that these estimates corroborate that the ionization of
the Cone is produced by S Mon. Similar conclusions have been reached before
by Schwartz et al. (1985) and Schmidt (1974).

There are two relatively bright (about 20 mJy at 6 cm; Girart et al. 2002)
sources about 8$'$ to the NW of the Cone Nebula.
These sources are known to have non thermal spectra 
(Schwartz et al. 1985) and most probably are background galaxies.
This result suggests that the non thermal jet may be associated with
one of the galaxies of a extragalactic background cluster. 
Unfortunately, given the large obscuration of the region, there are no
optical counterparts to any of these sources to test further the
possible presence of a cluster. 


\section{Summary}

We searched for proper motions and flux variability in the NGC 2264 non-thermal
radio jet. We found no proper motions larger (at a 3-$\sigma$ upper limit)
than 1.6 mas yr$^{-1}$. This stringent upper limit appears to rule out
a galactic location for the jet, either in the case of a thermal jet emanating
from a young star or a relativistic microquasar.
We found flux variations only 
in one knot of the jet, that we identify as the core
of the source. 

The high degree of linear polarization in the jet and its spatial structure are 
comparable with other cases of extragalactic jets. With this evidence, 
we believe that this object is an extragalactic jet seen in the
line of sight toward the NGC 2264
star forming region, even when the \sl a priori \rm probability of such 
coincidence is very low.

\begin{acknowledgements}

We thank an anonymous referee for many valuable suggestions.
LFR acknowledges the support
of DGAPA, UNAM, and of CONACyT (M\'exico).
This research has made use of the SIMBAD database, 
operated at CDS, Strasbourg, France.
The Digitized Sky Survey was produced at the Space Telescope Science 
Institute under U.S. Government grant NAG W-2166. The images of these surveys 
are based on photographic data obtained using the Oschin Schmidt Telescope on 
Palomar Mountain and the UK Schmidt Telescope. The plates were processed into 
the present compressed digital form with the permission of these institutions.

\end{acknowledgements}


\newcommand\rmaap   {RMA\&A~}
\newcommand\aujph   {Aust.~Jour.~Phys.~}
\newcommand\rmaacs   {RMA\&A Conf. Ser.~}




\end{document}